\documentclass[prd,a4paper,twocolumn,floatfix]{revtex4}
\usepackage{epsfig}
\begin{document}
\title{A new window on Strange Quark Matter as the ground state of
 strongly interacting matter}
\author{Vikram Soni}
\email{vsoni@del3.vsnl.net.in}
\affiliation{National Physical Laboratory, K.S. Krishnan Marg, New Delhi 110012, India}
\author{Dipankar Bhattacharya}
\email{dipankar@rri.res.in}
\affiliation{Raman Research Institute, Bangalore 560080, India}
\begin{abstract}
If strange quark matter is the true ground state of matter, it must have lower
energy than nuclear matter. Simultaneously, two-flavour quark matter must have
higher energy than nuclear matter, for otherwise the latter would convert to
the former. We show, using an effective chiral lagrangian,
that the existence of a new lower energy ground state for two-flavour quark 
matter, the pion condensate, shrinks the window for strange quark matter
to be the ground state of matter and sets new limits on the current
strange quark mass.
\end{abstract}
\maketitle
\section{\label{sec:intro}Introduction}
The hypothesis that the true ground state of baryonic matter may have a
roughly equal fraction of u,d and s quarks, termed strange quark
matter (SQM) is of recent origin \cite{ref1}. This is based on the fact that at
some density, when the down quark chemical potential is larger than the
strange quark mass, conversion to strange quarks can occur. This reduces
the energy density by having 3 (u, d and s) fermi seas instead of just 2 
(u, d), and can yield a state of energy lower than nuclear matter. It is also
possible to explain why such a state has escaped detection.

This involves at least two puzzles.
\begin{itemize}
\item[i)] Why does ordinary 2 flavour nuclear matter, the observed ground state
of baryonic matter, not decay into strange quark matter.

The answer  is that this decay is not like the radioactive decay
of unstable nuclei, as the nucleons cannot decay one by one as it is not
energetically favourable for the nucleon to change into a  $\Lambda$,
but only for the entire nuclear matter to transmute into strange quark
matter and this requires a high order of the flavour changing weak
interaction which renders the cross section to be exponentially  and
unobservably small.

\item[ii)] Why was this matter not created in the evolution of the universe?

This is due to the fact that as the universe cooled past a temperature
equivalent to the strange quark mass, strange quark matter was not
the chosen state of high entropy. Since the u and d quarks have
almost neglible masses at this scale, as the temperature dropped further
the strange quarks were Boltzmann suppressed leaving just the u and d
quarks, which as we know converted largely into nucleons. For details we refer
the reader to \cite{ref1,ref2,ref3}.
\end{itemize}

It is really quite remarkable that the ground state cannot be realised
easily! Only if we can produce high baryon density by compression
can SQM be realised - for example, in the interior of neutron stars.

We now turn to the theoretical underpinning of the case for SQM being
the potential ground state of matter.

We already know, empirically as well as theoretically, the ground
state energy of saturation nuclear matter - 930 MeV for the Fe$^{56}$ nuclei.
However, for calculating quark matter we take recourse to 
phenomenological models, which are pointers
but foundationally inadequate and here lies the uncertainty.

The usual ground state calculation for SQM treats the quarks as a free fermi
gas of current quarks. The volume in which these quarks live comes at a
cost of a constant energy density that provides `confinement.' It  is 
equivalently the same constant
value of negative pressure and hence is often called the bag pressure term.
This is a simple extension of the MIT bag philosophy, where the origin
of the constant energy density is the fact that quarks are confined. The bag
pressure sets the equilibrium or ground state energy density and the baryon
density. It can be fixed from the nucleon sector. Further structure can
be introduced by adding interaction between the quarks, e.g., one gluon
exchange. Such a phenomenological model has been used by Witten
\cite{ref1} and later by Farhi and Jaffe  \cite{ref2} and others
for SQM (see  \cite{ref3} for a review).

\section{Chiral Symmetry}

It is clear that such a model is phenomenological and does not, for example,
address the issue of the spontaneous breaking of chiral symmetry -- 
an essential feature of the strong interactions.

The quark matter in the bag is in a chirally  restored state. This means that
as  in the case of Superconductivity it costs energy to expel the chiral 
condensate which characterises the true vacuum state. Clearly, this will 
act just like the bag energy density/pressure. However, its value will be 
determined by the energy density of the chiral condensate. Such a term binds 
but does not confine. Confinement thus requires further input than just
a bag pressure.

All results for the SQM state will depend on the model that is
used to describe it and the ground state thereof. In a chiral model we find that
there is a plurality of ground states. Of these, we find that one particular
ground state has the property of chiral restoration at high density and
parallels the MIT bag state used in most previous estimates, where the ground
state is a fermi sea of current quarks (chirally restored quark matter or
CRQM) with the bag pressure provided by absence of the chiral condensate.
This regime sets the connection between the paramaters of the chiral
model and the MIT bag model used in  \cite{ref1,ref2,ref3}.

Unlike for the MIT bag case where the bag pressure is a parameter, in
our formulation, it is  the chiral condensate energy and is given in terms of
the parameters of low energy phenomenology - the pion decay constant, $f_\pi$, 
which is precisely known and the scalar coupling or the $\sigma$ mass, 
which is rather poorly `known'.

There are, however, other ground
states for this model, in which the pattern of symmetry breaking is different 
at high density, for example, the pion condensed (PC) ground state in which 
the chiral symmetry is still spontaneously broken at high density. Such a 
state has lower  ground state energy than the former and thus needs to be 
considered in the description of quark matter. As we show it is found to 
influence the regime of existence of SQM importantly.

\subsection{Effective chiral lagrangian}
We consider this issue in the framework of an intermediate Chiral Lagrangian
that has  chiral SSB. Such an effective
Lagrangian has quarks, gluons and a chiral multiplet of $[\vec\pi ,\sigma ]$
that flavor couples only to the quarks. For $SU(2)_L \times SU(2)_R$ chiral
symmetry, we have
\begin{widetext}
\begin{equation}
  L = - \frac{1}{4} G^a_{\mu \nu} G^a_{\mu \nu}   \\
 - \sum {\overline{\psi}} \left( D + g_y(\sigma +  \\
i\gamma_5 \vec \tau \vec \pi)\right) \psi            \\
- \frac{1}{2} (\partial                                \\
\mu \sigma)^2 - \frac{1}{2} (\partial \mu \vec \pi)^2    \\
-  \frac{\lambda^2}{4}(\sigma^2 + \vec \pi^2 - (f_\pi)^2)^2 \\
\end{equation}
\end{widetext}
The masses of the scalar (PS) and fermions follow on the
minimization of the potentials above. This minimization yields
\begin{equation}
\qquad  <\sigma>^2 = f_\pi^2
\end{equation}
where $f_\pi$ is the pion decay constant.
It follows that
\begin{equation}
\qquad m^2_{\sigma} = 2\lambda^2( f_\pi)^2
\qquad m_q = m= g <\sigma> = g f_\pi
\end{equation}
This theory is an extension of QCD by additionally coupling the quarks
to a chiral multiplet, $(\vec\pi$ and $\sigma)$ 
\cite{ref4,ref5,ref6}.

This Lagrangian has produced some interesting physics at the mean
field level \cite{ref6,ref7}:
\begin{enumerate}
\item It provides a quark soliton model for the nucleon in which the nucleon
 is realized as a soliton with quarks being bound in a skyrmion
 configuration for the chiral field expectation values
\cite{ref5,ref6}.
\item Such a model gives a natural explanation for the `proton spin puzzle'.
 This is because the quarks in the background fields are in a spin, isospin
 singlet state in which the quark spin operator averages to zero. On the
 collective quantization of this soliton to give states of good spin and
 isospin the quark spin operator acquires a small non zero contribution
 \cite{ref8}.
\item Such a Lagrangian also seems to naturally produce the Gottfried sum
  rule \cite{ref9}.
\item Such a nucleon can also yield from first principles (but with some
  drastic QCD evolution), structure functions for  the nucleon which are close
  to the experimental ones \cite{ref10}.
\item In a finite temperature field theory such an effective Lagrangian also
  yields screening masses that match with those of a finite temperature QCD
  simulation with dynamical quarks \cite{ref11}.
\item This Lagrangian also gives a consistent equation of state for strongly
 interacting matter at all density  \cite{ref12,ref6}.
\end{enumerate}
We shall first briefly establish the parameters of the above effective
Lagrangian and the specific connection with the MIT bag model of confinement
used in previous treatments of SQM.

As already pointed out above, the nucleon in this model is realised
as a  soliton in a chiral symmetry broken background with quark bound states
\cite{ref5,ref6,ref7}. This sets the value of the yukawa coupling, $g$, required to fit the
nucleon mass in a Mean Field Theory (MFT) treatment to be, $g = 5.4$.

For the nucleon the dependence on the scalar coupling, $\lambda$, is
marginal as long as it is not too small. Further, in MFT, the QCD
coupling does not play a role; only if 1 gluon exchange is included
does the QCD coupling enter.

There are no other parameters except $f_\pi$, the pion decay constant
which is set to 93 MeV.

\section{Preliminaries for SQM}
The connection to MIT bag description of quark matter is set as follows.
The last term in the above lagrangian, the potential functional,
\[ \frac{\lambda^2}{4} (\sigma^2 + \vec \pi^2 - (f_\pi)^2)^2  \]
is minimized by the VEV's
\[ (<\sigma> = f_\pi, <\vec \pi>= 0 )\]
and is equal to zero at the minimum.

In MFT at high density (as we shall see), when chiral symmetry is restored,
$(<\sigma>=0$, $<\vec\pi>=0)$, this term
reduces to  a constant energy density term equal to
\[ \frac{\lambda^2}{4} (f_\pi)^4 \]
Besides, due to chiral symmetry restoration the constituent mass of the
quarks also vanishes, leaving free massless quarks. This reduced lagrangian
for high density is no different from MIT bag quark matter with
\[ B   =\frac{\lambda^2}{4} (f_\pi)^4  \]
This completes the identification of the bag pressure term in this model.
It shows that bag pressure is automatically generated by chiral
restoration and is controlled simply by the scalar coupling or equivalently
the sigma mass.

We first briefly describe the logical basis for the investigation of
the QM ground states vis a vis the usual nuclear matter ground state
at saturation density. Here we follow Farhi and Jaffe \cite{ref2}. 
\begin{enumerate}
\item We fix coordinates by noting that SQM can be the true ground state only if
  its energy per baryon, $E_B$, is lower than the lowest value found in nuclei,
  930 MeV for iron, as done by Farhi and Jaffe, \cite{ref2}.
\item We calculate the 2 flavour quark matter ground states and fix a lower
  bound for the only free parameter in our lagrangian, the scalar coupling; or
  equivalently, we get a lower bound on the chiral condensate pressure (or bag 
  pressure)
  from the condition that the 2 flavour quark matter state must have higher
  $E_B$ than nuclear matter - otherwise nuclear matter would
  be unstable to conversion to the 2 flavour QM. As pointed out in
  \cite{ref2} this condition is that bulk 2 flavour quark matter must
  have $E_B >934$  MeV.
\item We calculate the SQM with the parameters established in  2 above and see
  if for SQM, $E_B$ is smaller than that given in 1, above. If this is the 
  case, and as $E_B$ increases monotonically with the scalar coupling (or
  the chiral condensate pressure), we get an upper bound on the chiral
  condensate pressure (or bag pressure), when $E_B$
  crosses beyond 930 MeV. SQM can then exist, as the true ground state, in
  this interval between the two bounds.
\end{enumerate} 

\section{Two Flavour Quark Matter}
We shall now consider in Mean Field Theory the phases of 2 flavour quark
matter in the $ SU(2)_L \times SU(2)_R $  chiral model above.
We shall then extend the model to 3 flavours (u, d, s) to describe SQM.

\subsection{The space uniform phase}
We now turn to the phase in which the pattern of symmetry breaking is 
such that the expectation values of the meson fields are uniform. At zero
density they are just the VEVs. 
\begin{eqnarray}
             <\sigma> &=& f_{\pi} \\
             <\vec \pi> &=& 0 
\end{eqnarray}
For arbitrary density we allow the expectation value to change in
magnitude, as it becomes a variational parameter that is determined
by energy minimization at each density.
\begin{eqnarray}
            <\sigma> &=& F \\
            <\vec\pi> &=& 0
\end{eqnarray}
Such a pattern of symmetry breaking simply provides a constituent mass to the 
quark $ m = g<\sigma> = gF $ and the quarks are in plane wave states
as opposed to the bound states in the nucleonic phase \cite{ref12}.

The mean field description of this phase is simple.

The energy density
\begin{equation}
           \epsilon_{\rho} = \Sigma_{u,d} \frac{1}{(2\pi)^3} \gamma
           \int d^3k \sqrt{m^2 + k^2}  + \frac{\lambda^2}{4} (<\sigma^2>
           -f_{\pi}^2)^2
\end{equation}
where $m = g<\sigma> = gF$  and the degeneracy $\gamma = 6$. We shall 
use $g = 5.4$ as determined
from fixing the nucleon mass in this model at 938 MeV \cite{ref5,ref6}.
The integral above runs up to the `u' and `d' fermi momenta.

For neutron matter (without $\beta$  equilibrium) we have the relations 
\begin{eqnarray}
    k^f_u &=& (\pi^2 n_u)^{\frac{1}{3}} = (\pi^2\rho_B)^{\frac{1}{3}} \\
    k^f_d &=& (2 \pi^2\rho_B)^{\frac{1}{3}} \\
    E_B  &=& \frac{\epsilon_{\rho}}{\rho_B} 
\end{eqnarray}
where $\rho_B$ is the baryon density.
At any density the ground state follows from minimising the free energy 
w.r.t. $<\sigma> = F$.

As shown in the figures of \cite{ref12,ref13}, this phase begins at
zero $\rho_B$, with $E_B = 3gf_\pi$, which then falls till chiral 
restoration
occurs at some $\rho_X$. After this, as density is increased the $E_B$
continues
to drop and goes to a minimum and then starts rising corresponding to a
massless quark fermi gas.

In the chirally restored phase the EOS is very simple and parallels the
MIT bag description of \cite{ref3}
\begin{eqnarray}
     \rho_B &>& \rho_X\\ 
 \epsilon_{\rho}
            &=& \frac{3}{4\pi^2} {\pi^2\rho_B}^\frac{4}{3}\alpha
                          + \frac{\lambda^2}{4} f_{\pi}^4 
\end{eqnarray}
 The last term above is just the bag energy density, and
\begin{equation}
 \alpha = (1 + 2^{\frac{4}{3}})  
\end{equation}
This phase has two features, a chiral restoration at $\rho_X$ followed,
with increasing density, by an absolute minimum in $E_B$, at
     $\rho_C >\rho_X$
         
Since $E_B$ decreases monotonically with density till the chiral
restoration density, $\rho_X$, and then  continues to decrease till
the minimum is reached at $\rho_C$, this implies that the density regime
till $\rho_C$ is unstable and has negative pressure. This has been 
recently 
conjectured \cite{ref14} as the density at which self-bound droplets 
of quarks
form, which may  be related to nucleons.  Further, since at this
density chiral symmetry is restored, these `nucleons' will be like those 
in the MIT bag model in which chiral symmetry is unbroken inside the
nucleons.

We would like to clarify this issue. 

From the comparison of this phase with the nucleon and nucleonic `phase'
arising from the same model (see \cite{ref6,ref12}), 
it is clear that the nucleonic phase
is always of lower energy than the uniform phase above, upto a density 
of roughly 3 times the nuclear density,
which is above the chiral restoration density in the uniform phase.
Further, the minimum in the nucleonic phase occurs very much below the 
minimum in the uniform phase.

The chiral restoration density in the uniform phase is thus not of any
physical interest as matter will always be in the lower energy nucleonic 
phase and so the identification of nucleon as a quark droplet at the
density at which the minimum occurs in the uniform phase is not viable.
Clearly, the nucleon is a quark soliton of mass M =938 MeV and falls at
the zero density limit in the  nucleonic phase.

\subsection{The Pion Condensed phase}
Here we shall consider another realization
of the expectation value of  $<\sigma>$ and $<\vec\pi>$ corresponding
to pion condensation. This phenomenon was first considered in the context
of nuclear matter.

Such a phenomenon also occurs with our quark based chiral $\sigma$
model and was first considered at the Mean Field Level by Kutschera
and Broniowski in an important paper  \cite{ref13}. Working in the 
chiral limit they found that the pion condensed
state has lower energy than the uniform symmetry breaking state (phase 2)
we have just considered for all density. This is expected, as the ansatz 
for the PC phase is more general than for phase 2.

The expectation values now carry a particular space dependence
\begin{eqnarray}
      <\sigma> &=& F  \cos{(\vec q. \vec r)} \\
      <\pi_3> & =&  F  \sin{(\vec q. \vec r)} \\
      <\pi_1> &=& 0 \\
      <\pi_2> &=& 0
\end{eqnarray}
Note, when $|\vec q |$ goes to zero, we recover the uniform phase 2.

The Dirac Equation in this background is solved in \cite{ref13}
and reduces to
\begin{equation}
     H \chi(k) = (\vec \alpha . \vec k - \frac{1}{2} \vec q . \vec \alpha \gamma_5 \tau_3 + \beta m) \chi (k) = E(k)\chi(k)   
\end{equation}
where $m=gF $

The extra term has been recast in terms of the relativistic spin
operator, $\vec \alpha \gamma_5$ .  It is evident that if spin is parallel to $ \vec q$ and
$ \tau_3 = +1$
(up quark) this term is negative and if $\tau_3 = -1 $ (down quark) it is positive.
For spin antiparallel to $\vec q$ the signs for $\tau_3 = +1$ and $-1$  are
reversed.

The spectrum for the hamiltonian is the quasi particle spectrum and
can be found to be
\begin{eqnarray}
      E_{(-)}(k) &=& \sqrt{ m^2 + k^2 +\frac{1}{4}q^2 -\sqrt{m^2q^2 +
                          (\vec q.\vec k )^2}} \\
 E_{(+)}(k) &=&  \sqrt{m^2 + k^2 + \frac{1}{4}q^2 + \sqrt{m^2q^2 +
                           (\vec q.\vec k)^2}} 
\end{eqnarray}
The lower energy eigenvalue $ E_{(-)}$ has spin along $ \vec q$ for
$ \tau_3 =1$, or has spin  opposite to $\vec q$ for $\tau_3 =-1$. 
The higher energy eigenvalue
$E_{(+)}$ has spin along $\vec q$ and $\tau_3 = -1$, or has spin
opposite to $\vec q $ and $ \tau_3 = +1$.

In  this  background the fermi sea in no longer degenerate in spin
but gets polarized into the  states above. The quasi particles are,
however, states of isospin.
We describe matter at a given fermi energy  of u and d quarks
set by their respective densities and  by charge neutrality
(corresponding to  say neutron like matter).

First we fill up all the  lower energy, $E_{(-)}(k)$, states and
then we have a gap and start filling up the  $E_{(+)}(k)$ states
till we get to $E_{F}^i$ ,the fermi energy corresponding to a given
density for each flavour.
\begin{widetext}
\begin{eqnarray}
  \rho_{i} &=&\frac{1}{(2\pi)^3}\gamma (\int d^3k\Theta( E_F^i- E_{(-)}(k)) +
            \int d^3k \Theta( E_F^i- E_{(+)}(k))) \\
  \rho_{B} &=& (\rho_u + \rho_d)/3 \\
  \epsilon_i &=& \frac{1}{(2\pi)^3}\gamma (\int d^3k E_{(-)}(k)
  \Theta( E_F^i- E_{(-)}(k)) +
            \int d^3k E_{(+)}(k) \Theta( E_F^i- E_{(+)}(k))) \\
  \epsilon_\rho &=& \epsilon_u  +  \epsilon_d + \frac{1}{2} F^2 q^2 + \frac{\lambda^2}{4}( F^2 - f_{\pi}^2)^2
\end{eqnarray}
\end{widetext}
we can now write down the equation of state as in Ref.~13.  It is found 
that the PC state is always lower in energy then the uniform phase 2.
For the explicit numbers and figures we refer the reader to \cite{ref13}.

We briefly remark on some features of this phase:
\begin{enumerate}
\item The 2 flavour PC state is quite different from the uniform phase: unlike 
   the 2 flavour CRQM states
   considered in \cite{ref2}, it cannot be recovered from 3 flavour CRQM
   by taking the strange quark mass to infinity. As we shall see in the next
   sections, this gives a new feature - a maximum strange current quark mass
   for SQM to be the true ground state.
\item The reason that the PC phase has energy lower than the uniform $<\sigma>$
   condensate is perhaps best understood in the language of quarks and 
   anti quarks. To make a condenste a quark and antiquark must make a
   bound state and condense.  For a uniform $<\sigma>$ condensate the $q$ and 
   $ \bar{q}$ must have equal and opposite momentum. Therefore, as the quark 
   density 
   goes up the system can only couple a quark with $k > k_f$ and 
   a $ \bar{q}$ with the opposite momentum . This costs much energy so the
   condensate can only occur if $k_f$ is small, at low density.                                    
   On the other hand, the pion condensed state is not uniform. So at finite
   density, if we take a quark with $k = k_f$ the $\bar{q}$ can have momentum
   $k =  |\vec k_f -\vec q |$, which is a much smaller energy cost.
\item Since the pion condensate is a chirally broken phase,
   the chiral restoration shifts from very low density in the uniform phase
   to very high density: $\sim 10 \rho_{nuc}$.  This is a signature of this 
   phase.
\item Since this phase is always lower in energy than the uniform phase  
  we go directly from the nucleonic phase  to the PC phase  completely 
  bypassing the uniform phase, and thus all the interesting features and 
  conjectures for the uniform phase are never realized.
\item Another feature of this $\vec\pi$ condensate is that since we have a spin
  isospin polarization we can get a net magnetic moment in the ground state.
\end{enumerate}

\section{The three flavour state }
The extension of the above to 3 flavours or SU(3) chiral symmetry needs some 
clarification.

The generalized Dirac Equation for the SU(3) case is considerably more
complicated and involves a singlet $ \xi_0$  and an SU(3) octet $ \xi_a$
of scalar fields and a  singlet $ \phi_0$  and an SU(3) octet $ \phi_a$
of pseudoscalar fields, that interact with the quarks as shown in 
\cite{ref15}.
\begin{widetext}
\begin{equation}
   H \psi(k) = (-i\vec \alpha .\vec \partial -
               g \beta(\sqrt{2/3}( \xi_0 +i\phi_0 \gamma_5) +
               \lambda^a( \xi_a +i\phi_a \gamma_5))) \psi  =  E\psi
\end{equation}
\end{widetext}
In the chiral limit, the spontaneous symmetry breaking pattern is not unique.
We choose the pattern in which the $SU(3)_L \times SU(3)_R$ chiral
symmetry breaks down to a vector SU(3). For the uniform case, we have
\begin{eqnarray}
        <\xi_0> &=& \sqrt{3/2} f_{\pi}\\
        <\xi_a> &=& 0 \\
       <\phi_0> &=& 0  \\
       <\phi_a> &=& 0 
\end{eqnarray}
This gives a constituent mass  $m = gf_\pi$ for all (u, d and s)
quarks. The explicit symmetry breaking strange quark mass term with
mass $ m_s$, is then added to $H$. The strange quark mass, $M_s$, then 
turns out to be the sum of the constituent and  explicit mass,
$M_s = gf_\pi + m_s$.

\subsection{The three flavour Pion Condensed phase}

For describing strange quark matter we use the 3 flavour Pion Condensed state.
This is  a more versatile state than the one used in \cite{ref2}
(3 flavour CRQM),  the latter being in a subset of the former.

Next, we formulate the symmetry breaking in the presence
of the pion condensate. This is given as follows:
\begin{eqnarray}
     <\xi_0> &=& \sqrt{3/2} F(1 + 2\cos{(\vec q.\vec r)})/3 \\
     <\xi_8> &=&- \sqrt{3} F(1 - \cos{(\vec q.\vec r)})/3   \\
    <\phi_0> &=& 0   \\
    <\phi_3> &=&  F (\sin{(\vec q.\vec r)})   
\end{eqnarray}
and all other fields have expectation value zero.

This gives exactly the PC hamiltonian equation for the u,d sector
and yields a simple mass relation above for the strange quark:
$ M_s = gF + m_s $;
when $q=0$ and $m_s=0$ we recover the chiral limit above.

We may now simply add the 2 flavour PC results for the
energy density and density derived above to the strange quark
energy density which arises from the single particle relation,
\[ E_s = \sqrt{ M_s^2 + k^2} \]

The strange quark energy density is given by  Baym \cite{ref16}, eq.~(8.20)
\begin{equation}
    \epsilon_s = \frac{3}{\pi^2 8}M_s^4 (x_s n_s( 2 x_s^2 + 1) -
                 ln(x_s + n_s))
\end{equation}
where $x_s = k_s^f/M_s$  and    $n_s = \sqrt{ 1 + x_s^2}$;
$k_s^f$ is the fermi momentum for the strange quarks.

The total energy density of the quarks for the 3 flavour PC is given by
\begin{equation}
 \epsilon_\rho = \epsilon_u  +  \epsilon_d + \epsilon_s + 
                 \frac{1}{2} F^2 q^2  + 
		 \frac{\lambda_1^2}{4}(F^2 - f_\pi^2)^2 
\end{equation}
From the effective potential given in \cite{ref15} for the SU(3) case,
there is an extra
factor of 3/2 that multiplies the last term. This can be absorbed, as we have
done, by a redefinition: $ \lambda_1  =  A \lambda $, where $A = \sqrt{3/2}$.

\subsection {$\beta$ equilibrium in the PC phase}     
We have the following general chemical potential relations
for quark matter
\begin{eqnarray}
        E^u_F &=& \mu_ u \\
        E^d_F &=& \mu_ d =  \mu_ s \\ 
       \mu_ e &=& \mu_ d - \mu_ u \\ 
          n_e &=& \frac{\mu_e^3}{3\pi^2} 
\end{eqnarray}
The charge neutrality condition, below, further reduces the number of 
independent chemical potentials to one.
\begin{equation}
      \frac{2 n_u (\mu_u, q , F) - n_d (\mu_d, q ,F)
          - n_s (\mu_s)}{3} -n_e  =  0 
\end{equation}
The baryon density is
\begin{eqnarray}
     \rho_{B} &=& \frac{n_u (\mu_u, q , F) + n_d (\mu_d, q ,F) 
                  + n_s (\mu_s)}{3} \\
          n_s &=& (k_s^f)^3/(\pi^2)
\end{eqnarray}
For matter in $\beta$ equilibrium we need to add the electron energy density
to the quark energy density above
\[ \epsilon_e =  (1/4 \pi^2) \mu_e^4 \]
The total energy density is
\[ \epsilon  =  \epsilon_\rho + \epsilon_e \]
The energy per baryon, $E_B = \epsilon / \rho_B$, then follows.

For the pion condensed state, the ground state energy and the baryon density
depend on the variational parameters, the order parameter or the
expectation value, $F = \sqrt{ <\vec \pi>^2 + < \sigma>^2 }$ and the
condensate momentum, $|\vec q|$. To define the free energy at a fixed
baryon density then requires some care. However, we are only interested
in the absolute mininmum of the energy per baryon, $E_B$, for all density.
We can then simply minimize $E_B$ with respect the variational parameters,
$F$ and $|\vec q|$ and the one independent chemical potential to get the
absolute minimum.
Note that this is for a given a value of
\begin{equation}
  B =  \lambda_1^2 ( f_\pi)^4 /4 
\end{equation}

\section{Results for the mean field theory 3 flavour PC state (PCSQM)}

 \subsection{PCSQM without one gluon exchange}

The new window for SQM is established thus:

We maintain the minimum permissible limit on $E_B$ for 2 flavour quark 
matter to be
934 MeV. Since the PC is a lower energy state than the chirally restored
QM (CRQM) considered in Ref.~2 we find that the lower bound on
$B^{1/4}$ goes up, closing the window on SQM. The results are given in 
table~\ref{table1}.
This lower bound for 2 flavour PC is found to be
\[ B^{1/4}_<  = 148 \mbox{\rm ~MeV} \]
whereas, for the case of 2 Flavour CRQM considered by Farhi and Jaffe, it was
\[ B^{1/4}_<  = 145 \mbox{\rm ~MeV} \]

We then calculate the minimum $E_B$ for 3 flavour QM. For this we use our
generalised PCSQM as the state. It is important to note that the rather
particular 3 flavour CRQM lies within its variational reach. This provides us
with the upper bound on the bag pressure $B_>$.
\begin{table}
\caption[x]{\label{table1}
Ground states with energy of 930 MeV/nucleon for 3-flavour quark
matter with 2-flavour Pion-condensed state. $m_s$ is the assumed strange quark
mass, $B$ stands for Bag Pressure, $\mu_{\rm u}$ the u-quark chemical
potential, $n_{\rm s}/n_{\rm u}$ the ratio of the density of strange
quarks to that of u-quarks and $< \sigma >$ is the expectation value of the
$\sigma$ field. These results are without 1-gluon exchange}.
\begin{tabular}{rcccr}
\hline
\multicolumn{1}{c}{$m_s$} & $B^{1/4}$ & $\mu_{\rm u}$ & $n_{\rm s}/n_{\rm u}$ &
$< \sigma >$ \\
\multicolumn{1}{c}{(MeV)} & (MeV) & (MeV) &  & (MeV) \\
\hline
  0 & 162.7 & 309 & 0.96 &  9.55 \\
 50 & 161.4 & 307 & 0.96 &  1.45 \\
100 & 158.9 & 304 & 0.92 &  0.01 \\
150 & 154.9 & 298 & 0.80 &  0.01 \\
200 & 150.6 & 290 & 0.63 &  0.01 \\
250 & 147.7 & 260 & 0.00 & 26.31 \\
\hline
\end{tabular}
\end{table}
The maximum upper bound on $B$ naturally occurs for the $m_s = 0$ case
and is found to be almost the same as in Ref.~2.
\[ B^{1/4}_>  = 162.5 \mbox{\rm ~MeV} \]
This is because the minimum occurs in the 3 flavour CRQM state, as given
in \cite{ref2}, with $F = 0$.

Some new features compared to Ref.~2:

Whereas in \cite{ref2}, the 2 Flavour CRQM threshold is virtually the
limit of $m_s \rightarrow \infty$ for 3 flavour CRQM, in our case it is
not, as our 2 flavour PC ground state is of lower energy and different. In
this case the 3 flavour CRQM becomes of higher energy than the 2 flavour PC
ground state at a finite $m_s$.

We thus find the limit
\[ m_s  < 250  \mbox{\rm ~MeV} \]
for SQM to exist as the ground state, simply from the constraint on 2 flavour
QM.

\begin{figure}
\epsfig{file=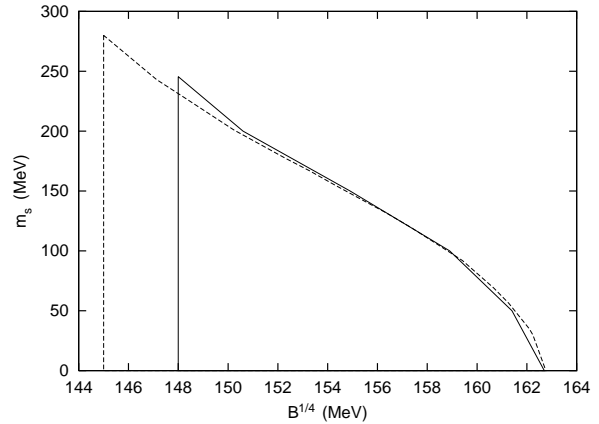,width=8cm}
\caption[x]{\label{figure1} Allowed region in the bag constant-strange quark 
          mass parameter
          plane for the Strange Quark Matter (SQM) ground state to be the 
	  absolute ground state of matter.  Gluon exchange interaction is not 
	  included.
          The solid line shows the allowed window, taking into account the
          Pion Condensed (PC) phase reported in this paper.  The region included
          within the curve is the allowed region.  The dashed line shows the
          result for Chirally Restored Quark Matter (CRQM) from ref.~\cite{ref2}.
          The vertical left boundary for either case represents $E_B = 934$~MeV
          in two-flavour matter and the curved line reresents $E_B = 930$~MeV
          in 3-flavour matter. The curve for CRQM shown here is a linear 
          interpolation to 930~MeV from the results for 919~MeV and 939~MeV 
          presented in ref.~\cite{ref2}.}
\end{figure}

We also find that for this limiting $m_s$, the absolute minimum of
$E_B$ is the 2 flavour PC state with zero strange quark density. But, for
masses somewhat below the limiting mass (with the condition that SQM 
be the actual ground state), the absolute minimum occurs in the 3 flavour 
CRQM state as given in \cite{ref2}, with $F=0$. A summary of the results 
appears in Fig.~\ref{figure1}.

\subsection{Results for the three flavour Pion Condensed phase (with 1 gluon 
exchange interaction and $\alpha_{QCD} = 0.6$)}

We next consider the case in which 1 gluon exchange interaction is included,
following Farhi and Jaffe. This is with a view to estimating the effects of
including such `perturbative' interactions. No attempt will be made at rigour.
Our approximate scheme for this case is best regarded as an estimate.

One reason it is difficult to do an analytic calculation of the
interaction energy for the  PC  is that the
quark propagators in the presence of the pion condensate \cite{ref13} are
far more complicated than in the case of free fermi sea quarks.

We first note that in the limit of the condensate, $F \rightarrow 0$,
all the PC results go smoothly to the free fermi sea results.
We find that the condensate expectation values $F$ in the regime of interest
to us are such that the value of $F$ at the minimum makes $m = gF$ fall below
the relevant quark chemical potentials. So, as a first approximation, we
use the free fermi sea results for the given chemical potential and mass.

The contribution from the 1 gluon exchange interaction to the free energy
at given density or the Thermodynamic Potential (TP), including
renormalization group corrections, (a), is given in \cite{ref2}.
The expression for the 1 gluon exchange or interaction energy, for free fermi
seas of quarks, (b), is given in Baym \cite{ref16} (eq.~8.20). 
Though
we expect the interaction contribution (in the absence of renormalization
group corrections) to both (a) and (b) to be the same, we find that the
two expressions are different. Results for these two cases are summarized
in table~\ref{table2}.
\begin{table*}
\caption[x]{\label{table2} Same as table~\ref{table1}, but interaction 
            energies due to 1 gluon exchange included.}
\begin{tabular}{r|cccr|ccrr}
\hline
 & \multicolumn{4}{c|}{(a) Farhi-Jaffe model} & \multicolumn{4}{c}{(b) Baym model} \\
\multicolumn{1}{c|}{$m_s$} & $B^{1/4}$ & $\mu_{\rm u}$ & $n_{\rm s}/n_{\rm u}$ &
$< \sigma >$ &
$B^{1/4}$ & $\mu_{\rm u}$ & $n_{\rm s}/n_{\rm u}$ & $< \sigma >$ \\
\multicolumn{1}{c|}{(MeV)} & (MeV) & (MeV) &  & (MeV) & (MeV) & (MeV) &  & (MeV) \\
\hline
  0 & 150.3 & 274 & 0.83 & 22.08   &  166.1 & 302 & 0.69 & 36.53 \\ 
 50 & 147.7 & 274 & 0.96 &  0.33   &  161.6 & 292 & 0.54 & 34.45 \\
100 & 144.6 & 280 & 0.89 &  0.01   &  157.2 & 282 & 0.28 & 35.54 \\
150 & 140.8 & 244 & 0.00 & 32.91   &  155.8 & 272 & 0.00 & 40.67 \\
200 & 140.8 & 244 & 0.00 & 32.91   &  155.8 & 272 & 0.00 & 40.68 \\
250 & 140.8 & 244 & 0.00 & 32.91   &  155.8 & 272 & 0.00 & 40.67 \\
\hline
\end{tabular}
\end{table*}

We briefly give a comparison with the results of the case without the 1
gluon exchange interaction. In our notation the results for (a) are given first (without brackets) and the
results for (b) appear alongside in brackets.

The lower bound for 2 flavour PC is found to be
\[ B^{1/4}_< =  141.5 \; (156.7) \mbox{\rm ~MeV} \]
For (a) this is down, from the case without 1 gluon exchange, where it was 
148 MeV.
This indicates that gluon exchange is repulsive, even with the constituent quark
mass generated by the condensate, whereas for (b) it is up and the 1 gluon exchange is attractive.

 We note that for the case of 2 Flavour CRQM (with gluon exchange) considered
by Farhi and Jaffe, we find
\[ B^{1/4}_<  = 132 \mbox{\rm ~MeV} \]
is even more down, from the case without 1 gluon exchange, where
it was 145 MeV.

This clearly shows that the effect of gluon exchange is more repulsive for this
case as the quarks are massless as opposed to the PC case when they have
a mass, $m = gF$.
                 
We note that our value, $B^{1/4}_< = 132$~MeV, is more than that given in the 
figure 1(c) in \cite{ref2}. This is due to the difference in the way energy 
density and density are defined for us and in \cite{ref2}. They begin with the 
TP to order $\alpha$, derive a density which includes interaction to order 
$\alpha$, and use this density to define the energy density.
The difference between our case and theirs is $O(\alpha^2)$.

The maximum upper bound on $B$ naturally occurs for the $m_s = 0$ case
\[ B^{1/4}_>  =   150.3 \; (166.1) \mbox{\rm ~MeV} \]
Interestingly, in this case the minimum in $E_B$ comes from a {\em new ground 
state} and is genuinely different.
It does not occur either in the 2 flavour PC state  nor does it occur in the
3 flavour CRQM state as given in Farhi and Jaffe, with $F=0$,  as was the 
case in the absence of 1 gluon interaction. In this case the mininmum is lower
than either of the these states and comes from a true merger of the two;
it has a non zero value of, $F=22.1 \; (36.5)$~MeV, and also a ratio of the 
strange quark density to the u quark density of $0.83 \; (0.69)$.

For comparison for the 3 flavour CRQM state of \cite{ref2}
\[ B^{1/4}_>  = 144.5 MeV \]

We find the maximum allowed limit on $m_s$, with gluon exchange included, 
for both (a) and (b), moves down to
\[ m_s  <  150  \mbox{\rm ~MeV} \]
for SQM to be the absolute ground state.

We remark that the cases (a) and (b) show the same trends. The difference is 
that the allowed values of $B$ are shifted up in (b).

\begin{figure}
\epsfig{file=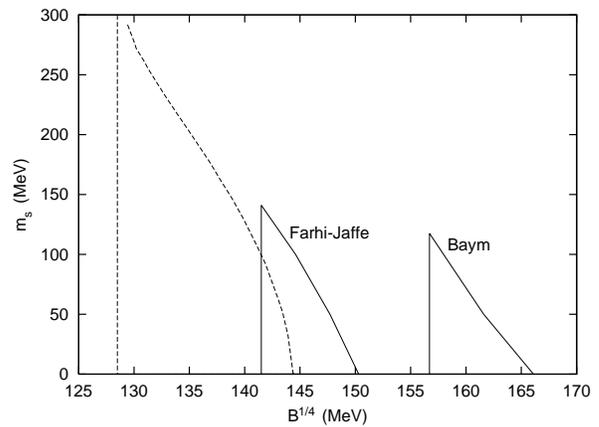,width=8cm}
\caption[x]{\label{figure2} Similar to Figure~\ref{figure1}, but with 
           1-gluon exchange included. The 
           exchange energy is computed using two alternative prescriptions, 
           by Farhi-Jaffe~\cite{ref2} and Baym~\cite{ref16}, as indicated.  
           The solid lines represent the constraints imposed by taking 
           into account the PC phase in these two cases.  The dashed line is 
           for CRQM, obtained from the data presented by Farhi and 
           Jaffe~\cite{ref2}.  $\alpha_{QCD}$ is set to 0.6. The PC phase
           places a strong constraint on the strange quark mass if SQM 
           should represent the absolute ground state of matter.}
\end{figure}

Some of these results are summarized in Fig.~\ref{figure2}.

\section{Conclusions}

We have found that the existence of a new and lower energy state of
2 flavour quark matter, the pion condensed state, has a significant effect
on the window of opportunity for SQM to be the true ground state of matter.

We work with an effective chiral lagrangian. Unlike the MIT bag case
\cite{ref2},  where the bag pressure is a parameter, in our formulation
it is the chiral condensate energy and is given in terms of the parameters
of low energy phenomenology - the pion decay constant, $f_\pi$, which
is precisely known and the scalar coupling or the $\sigma$ mass, which
is rather poorly known. However, it is of interest that SQM is related to
parameters of low energy phenomenology. It  requires more detailed work to
firm up this connection. Furthermore, we have found a new and interesting
constraint on the existence of SQM as the true ground state, that comes 
from the current mass of the strange quark.
 
Our finding is that
\begin{enumerate}
\item Without including 1 gluon exchange the new PC ground state
limits the bounds on the bag pressure $B$, allowing
\[ 148 \mbox{\rm ~MeV} <   B^{1/4} <  162.5 \mbox{\rm ~MeV} \]
instead of the result of \cite{ref2}
\[ 145 \mbox{\rm ~MeV} <   B^{1/4} <  162.5 \mbox{\rm ~MeV} \]
and cuts down the allowed pararmeter space of the explicit
or current strange quark mass to
\[  m_s  <  250  \mbox{\rm ~MeV} \] 
\item With $\alpha_{QCD} = 0.6$ and including 1 gluon exchange the new PC
ground state, with some simplifying approximations
strongly limits the bounds on the bag pressure, $B$ , allowing
\[ 141.5 (156.7) \mbox{\rm ~MeV} <   B^{1/4} <  150.3 (166.1) \mbox{\rm ~MeV} \]
instead of the result of \cite{ref2}:
\[ 128.5 \mbox{\rm ~MeV} <   B^{1/4} <  144.5 \mbox{\rm ~MeV} \]
and it further cuts down the allowed pararmeter space of the explicit 
strange quark mass to
\[  m_s  <  150  \mbox{\rm ~MeV} \]
This is a rather punishing constraint.
\end{enumerate}
These results are obtained in the 2 flavour chiral limit with some 
approximations
and also assuming that all the bag pressure comes from the chiral condensate.
Adding a confinement pressure will raise the energy, $E_B$, for SQM and may
shrink the window further.

This is, by no means, the last word on possible ground states even in
this model--for example, we may have a kaon condensate. However, we
shall not investigate this here.

\subsection*{Acknowledgements}
We would like to thank W. Broniowoski for sharing with us the program on the 
Pion Condensed state which is the mainstay of this paper. We genuinely thank
Judith Mcgovern for her ever willing help with the intricacies SU(3) chiral
model and more. We thank Bob Jaffe for very prompt clarifications on the 
expressions in their paper, which is the benchmark for this one. We further 
thank Mike Birse and  M. Kutschera. VS thanks the Raman Research Institute 
for its hospitality.


\newpage
\begin{thebibliography}{99}
\bibitem{ref1} E. Witten, Phys. Rev. D, 30, 272 ( 1984).
\bibitem{ref2} E. Farhi and R.L. Jaffe, Phys. Rev. D, 30, 2379 ( 1984).
\bibitem{ref3} C. Alcock and A. Olinto, Ann. Rev. Nucl. Part. Sci., 38, 161
(1988).
\bibitem{ref4} V. Soni, Mod. Phys. Lett. A, 11, 331 (1996),
  A. Manohar and H. Georgi, Nucl. Phys. B, 234, 203 (1984)
\bibitem{ref5}  S. Kahana, G. Ripka and V. Soni, Nucl. Phys. A, 415, 351 (1984),
M.C. Birse and M.K. Banerjee, Phys. Lett., 134B, 284, (1984).
\bibitem{ref6} V. Soni, `The nucleon and strongly interacting matter', Invited talk
at DAE Symposium in Nuclear Physics, Bombay, Dec 1992 and references therein
.
\bibitem{ref7} M.C. Birse, Soliton Models in Nuclear Physics, Progress  in
 Particle and Nuclear Physics, Vol. 25, 1 (1991), and references therein.
\bibitem{ref8} see for example, R. Johnson, N.W. Park, J. Schechter,
V. Soni and H. Weigel,
Phys. Rev. D42 , 2998 (1990), J. Stern and G. Clement, Mod Phys.
Lett, A3, 1657 (1988).
\bibitem{ref9} see for example, J. Stern and G. Clement, Phys. Lett. B, 264, 
426  (1991),
E.J. Eichten, I. Hinchcliffe and C. Quigg, Fermilab-Pub 91/272 - T.
\bibitem{ref10} D. Diakonov, V. Petrov, P. Pobylitsa, M. Polyakov and C. Weiss,
 Nucl. Phys. B, 480, 341 (1996), Phys. Rev. D, 56, 4069 (1997).
\bibitem{ref11} A. Gocksch, Phys. Rev. Lett.  67, 1701 (1991).
\bibitem{ref12} V. Soni, Phys. Lett., 152B, 231 (1985).
\bibitem{ref13} M. Kutschera, W. Broniowski and A. Kotlorz, Nucl. Phys. A, 516
  566 (1990) .
\bibitem{ref14} M. Alford, K. Rajgopal and F. Wilczek, Phys. Lett B,
 422, 247 (1998), Nucl. Phys. B, 537, 443 (1999).
\bibitem{ref15} J.A. McGovern and M.C. Birse, Nucl. Phys. A, 506, 367 (1990) .
\bibitem{ref16} G. Baym and S. Chin, Phys. Lett., 62B, 241 (1976),
 G. Baym in NORDITA lectures, `Neutron Stars and the Properties of Matter
 at high density', (1977).
\end{thebibliography}
\end{document}